\documentclass[prd,preprintnumbers,superscriptaddress,amsmath,amssymb,nofootinbib,twocolumn,floatfix]{revtex4-2}

\usepackage{graphicx}
\graphicspath{{figures/}}
\usepackage{hyperref}
\usepackage{booktabs}
\usepackage{amsmath,amssymb,mathtools}
\usepackage{xcolor}

\newcommand{\be}{\begin{equation}}
\newcommand{\ee}{\end{equation}}
\newcommand{\Basilic}{\textsc{Basilic}}

\begin{document}


\title{GW231123: lensed, or just flexible?}

\author{Iuliu Cuceu}
\email{icuceu@oca.eu}
\affiliation{Universit\'e C\^ote d'Azur, Observatoire de la C\^ote d'Azur, CNRS, Laboratoire Artemis, 06300 Nice, France}
\author{Marie Anne Bizouard}
\email{marieanne.bizouard@oca.eu}
\affiliation{Universit\'e C\^ote d'Azur, Observatoire de la C\^ote d'Azur, CNRS, Laboratoire Artemis, 06300 Nice, France}
\author{Nelson Christensen}
\email{nelson.christensen@oca.eu}
\affiliation{Universit\'e C\^ote d'Azur, Observatoire de la C\^ote d'Azur, CNRS, Laboratoire Artemis, 06300 Nice, France}

\date{\today}

\begin{abstract}
GW231123 is the most massive binary black hole merger observed, exhibiting strong waveform-model degeneracies and Bayes factors up to $10^5$ favoring gravitational lensing. Comparing four precessing waveform models with and without overlapping lensed images, we find that lensing collapses the evidence range between models from $4.35$ to $0.27$ decades. Simulated unlensed signals reproduce these Bayes factors and lower inferred spins by absorbing waveform mismatch and Occam penalties. Due to waveform systematics, the data cannot establish that GW231123 is lensed.

\end{abstract}

\maketitle

\section{Introduction}

The LIGO~\cite{LIGOScientific:2014pky}, Virgo~\cite{VIRGO:2014yos}, and KAGRA~\cite{KAGRA:2020tym} Collaboration (LVK) reported GW231123, a gravitational-wave signal observed by the two LIGO detectors on November 23, 2023~\cite{LIGOScientific:2025rsn}. It was interpreted as the merger of two black holes with a total mass of $190$--$265\,M_\odot$ and nearly extremal spins. The signal spans only $\sim5$ cycles, between 30 and 80~Hz~\cite{LIGOScientific:2025rsn,Cuceu:2025fzi}. Its inferred source properties depend strongly on the waveform model, although all models agree on a very massive binary whose heavier black hole spins rapidly~\cite{LIGOScientific:2025rsn}. Simulated signals, with or without Gaussian noise, show comparable discrepancies, which are traced to model disagreement at strong spin precession~\cite{Bini:2026kwz,Jan:2025zcm}.

GW231123 has motivated many interpretations beyond a quasicircular compact binary coalescence. Proposed origins for its masses and spins include hierarchical mergers~\cite{Li:2025pyo,Passenger:2025acb,Li:2025fnf}, primordial black holes~\cite{DeLuca:2025fln,Yuan:2025avq}, and Population III or chemically homogeneous progenitors~\cite{Angeloni:2026nmy,Popa:2025dpz}. Proposed descriptions of the signal include orbital eccentricity~\cite{Wang:2026lur} (see also Refs.~\cite{Jan:2025zcm,Chandra:2026voe}), two overlapping mergers~\cite{Hu:2025lhv}, short transient noise~\cite{Ray:2025rtt}, and birefringent propagation~\cite{Liu:2026wor}. A cosmic-string origin is decisively disfavored~\cite{Cuceu:2025fzi}. The ringdown has been analyzed separately~\cite{Wang:2025rvn,Siegel:2025xgb}. Several of these are favored over a quasicircular binary by $\log_{10}\mathcal{B}=1$--$4$, with values that depend on the waveform model.

Gravitational lensing has received much attention. The LVK lensing analysis found the largest lensing Bayes factor of its catalog for GW231123: for a point-mass lens, $\log_{10}\mathcal{B}^{\rm L}_{\rm U}=3.8$ with the IMRPhenomXPHM-SpinTaylor waveform model, but  $\log_{10}\mathcal{B}^{\rm L}_{\rm U}=0.1$ with IMRPhenomXO4a~\cite{LIGOScientific:2025cwb}. The authors noted that ``the favoring of lensing decreases for waveform models more favored by the data.'' More detailed lens models including wave-optics give $\log_{10}\mathcal{B}^{\rm L}_{\rm U}$ of up to 5.0~\cite{Goyal:2025eqo,Shan:2025dcd}. The probability that an unlensed signal produces such a Bayes factor is estimated at below 1\%~\cite{Goyal:2025eqo} to 2--8\%~\cite{Shan:2025dcd}, depending on the waveform model, and 8\% of over $7\times10^{4}$ simulated unlensed signals favor lensing in Ref.~\cite{Chan:2025kyu}. The prior probability of such lenses is discussed in Refs.~\cite{Cheung:2026pky,Vujeva:2026xfh}.

In this Letter, we compare lensed and unlensed hypotheses for GW231123 with four waveform models, using \Basilic~\cite{Cuceu:2026ggj}, a pipeline for Bayesian inference and model comparison of gravitational-wave transients. We then repeat the comparison on simulated signals drawn from the fitted hypotheses. For these, the hypothesis with the highest evidence is often not the true one: as found for waveform models~\cite{Bini:2026kwz}, the Occam penalty can favor the wrong model. Lensing Bayes factors inherit this, and the lensed model also absorbs the waveform models' inaccuracies. The lensing Bayes factor of GW231123 is dominated by these waveform systematics and cannot by itself establish lensing.

\section{Method}

We analyze the public LIGO Hanford and Livingston data of GW231123~\cite{gwoscGW231123,LIGOScientific:2025snk}, using the glitch-mitigated Hanford data of the LVK analysis~\cite{LIGOScientific:2025rsn} and noise power spectral densities estimated with \texttt{BayesWave}~\cite{Cornish:2014kda,Littenberg:2014oda}. The likelihood is the Gaussian-noise likelihood over 20--448~Hz~\cite{Christensen:2022bxb,Thrane:2018qnx}, on a $4$~s segment sampled at $1024$~Hz and high-pass filtered at 10~Hz. We marginalize it over the luminosity distance and the arrival time~\cite{Thrane:2018qnx}. We compute evidences with the nested sampler \texttt{dynesty}~\cite{2004AIPC..735..395S,10.1214/06-BA127,Speagle:2019ivv,sergey_koposov_2023_7600689} in \texttt{bilby}~\cite{Ashton:2018jfp,Romero-Shaw:2020owr}, with $n_{\rm live}=800$ live points, within the \texttt{Basilic} pipeline~\cite{Cuceu:2026ggj}.

The unlensed hypotheses are quasicircular precessing binary black holes. We use four waveform models with higher-order multipoles: the numerical-relativity surrogate NRSur7dq4~\cite{Varma:2019csw}, IMRPhenomXO4a (XO4a)~\cite{Thompson:2023ase}, IMRPhenomXPHM with SpinTaylor precession angles (XPHM-ST)~\cite{Pratten:2020ceb,Colleoni:2024knd}, and IMRPhenomXPNR (XPNR)~\cite{Hamilton:2025xru}. The binary parameters have the same prior in all eight compact-binary hypotheses. The masses are uniform in the detector-frame component masses, with chirp mass in $[70,180]\,M_\odot$, mass ratio in $[1/6,1]$, and total mass below $400\,M_\odot$. Spin magnitudes are uniform up to 0.99 with isotropic orientations, the luminosity distance is uniform in the source frame up to 10~Gpc, and the arrival time is uniform within $\pm0.2$~s of the trigger.

For each waveform model, the lensed hypothesis uses the phenomenological model of Ref.~\cite{Liu:2023ikc}. In the geometric-optics limit~\cite{Takahashi:2003ix}, the signal is a superposition of $K$ overlapping images of the same merger:
\be
\tilde h_{\rm L}(f)=\tilde h_{\rm U}(f)\sum_{j=1}^{K}\sqrt{\mu_{j1}}\;e^{-2\pi i f\Delta t_{j1}+i\pi n_j\,{\rm sgn}(f)}.
\label{eq:lens}
\ee
Here $\mu_{11}=1$, $\Delta t_{11}=0$, and $\tilde h(f)=\int h(t)\,e^{-2\pi i f t}\,dt$, so image $j$ arrives $\Delta t_{j1}>0$ after the first image. Its Morse index is $n_j\in\{0,1/2,1\}$~\cite{Dai:2017huk}. The relative magnifications $\mu_{j1}$ are log-uniform in $[0.01,100]$, the time delays are uniform in $[1,100]$~ms and ordered, and the Morse indices are uniform over their three values. We use $K=2$ for all models which adds four parameters. As $\mu_{j1}\to0$ the lensed model approaches the unlensed one, so its best fit is at least as good. Lensing of this type is partially degenerate with spin precession~\cite{Liu:2023emk}.

We also test two non-compact-binary models from the \Basilic\ catalog~\cite{Cuceu:2026ggj}. The first is a generic frequency-domain power law, a proxy for many burst-type signals, including cosmic strings~\cite{Cuceu:2025fzi}. The second is a half-damped sinusoid, a proxy for a short instrumental transient, or glitches. Each has seven parameters, including sky position and arrival time. The amplitudes are log-uniform up to $10^{-18}$, from $10^{-25}$ for the power law and $10^{-23}$ for the sinusoid. The power-law index is uniform in $[0.01,4]$ and its high-frequency cutoff $f_{\rm high}$ in $[25,448]$~Hz. The sinusoid frequency $f_0$ and damping time $\tau$ are uniform in $[20,448]$~Hz and $[1,100]$~ms.

For two hypotheses, the Bayes factor is the ratio of evidences, $\mathcal{B}_{12}=\mathcal{Z}_1/\mathcal{Z}_2$. We write $\mathcal{B}_{\rm s/n}$ for a signal hypothesis against Gaussian noise, $\mathcal{B}^{\rm L}_{\rm U}$ for the lensed against the unlensed version of a waveform model, and $\mathcal{B}^{A}_{B}$ for waveform model $A$ against $B$. We quote Bayes factors as $\log_{10}\mathcal{B}$, in decades, and their split into fit and information below in nats. We use equal prior odds and do not include the astrophysical probability of lensing~\cite{Cheung:2026pky,Vujeva:2026xfh}. Each Bayes factor decomposes exactly as~\cite{10.1214/06-BA127}
\be
\ln\mathcal{B}_{\rm s/n}=\langle\ln\mathcal{L}\rangle-H,\qquad H=\int P\ln\frac{P}{\pi}\,d\theta.
\label{eq:info}
\ee
Here $\langle\ln\mathcal{L}\rangle$ is the posterior mean of the log-likelihood ratio against noise, and $H$ is the Kullback--Leibler divergence of the posterior $P$ from the prior $\pi$, i.e., the information gained from the data. We evaluate $\langle\ln\mathcal{L}\rangle$ with the full likelihood, including the distance and arrival time, so that $H$ includes the information on these two parameters. The first term measures the fit, and the second is the Occam penalty. This split is not unique: Ref.~\cite{Bini:2026kwz} measures the fit with the maximum likelihood instead. $H$ also carries the statistical uncertainty of $\ln\mathcal{B}_{\rm s/n}$, $\sigma\simeq(H/n_{\rm live})^{1/2}$~\cite{10.1214/06-BA127}. A lensing hypothesis Bayes factor splits as $\ln\mathcal{B}^{\rm L}_{\rm U}=\Delta\langle\ln\mathcal{L}\rangle-\Delta H$. The lensing and binary priors are independent. The chain rule of the Kullback--Leibler divergence then splits the information of the lensed model as $H_{\rm L}=H_\ell+H_{{\rm s}|\ell}$. Here $H_\ell$ is the information on the lensing parameters, and $H_{{\rm s}|\ell}$ the information on the binary parameters given the lensing ones. The change in information is $\Delta H=H_\ell+\Delta H_{\rm s}$, with $\Delta H_{\rm s}=H_{{\rm s}|\ell}-H_{\rm U}$. We estimate $H_\ell$ from the posterior samples, exactly for the Morse indices and with a $k$-nearest-neighbor entropy estimator~\cite{Kozachenko1987} for the magnifications and time delays.

To interpret these Bayes factors, we analyze simulated signals of known origin. We draw their parameters at random from the posteriors of four hypotheses fitted to GW231123: unlensed and lensed XO4a, and unlensed and lensed XPHM-ST, the waveform models with the highest and the lowest unlensed evidence obtained for GW231123 as reported in Table~\ref{tab:event}. Of the 17 signals per hypothesis, twelve are added to Gaussian noise with the power spectral densities of the event: three parameter draws, each in the same four noise realizations. Four are added to the real detector data around the event, each in a different 4~s stretch centered 6 or 2~s before or after it, with the signal and its arrival-time prior shifted accordingly. These stretches contain the non-Gaussian features of the data, such as a short Hanford transient 4~s after the event, while the event itself contributes negligibly. The last signal is analyzed without noise, which isolates the effect of the models from that of the noise. The 68 signals have network optimal SNRs of 17.9--22.6. We analyze each one with the same four hypotheses, i.e., with both the waveform model that generated it and the other one, using the settings of the event analysis.

\section{Results}

\begin{table}
\caption{Hypotheses tested on GW231123: number of parameters $d$, Bayes factor against noise, evidence relative to the best hypothesis, median network matched-filter SNR $\rho$, posterior mean of the log-likelihood ratio $\langle\ln\mathcal{L}\rangle$, and information $H$, with $\ln\mathcal{B}_{\rm s/n}=\langle\ln\mathcal{L}\rangle-H$ [Eq.~\eqref{eq:info}]. The uncertainty on $H$ is the nested-sampling uncertainty of $\ln\mathcal{Z}$. A superscript L denotes the lensed hypothesis. The glitch model is the half-damped sinusoid. The hypotheses are ranked according to $\ln\mathcal{B}_{\rm s/n}$.}
\label{tab:event}
\renewcommand{\arraystretch}{1.12}
\setlength{\tabcolsep}{2.6pt}
\footnotesize
\begin{ruledtabular}
\begin{tabular}{lcccccc}
Hypothesis & $d$ & $\log_{10}\mathcal{B}_{\rm s/n}$ & $\Delta\log_{10}\mathcal{Z}$ & $\rho$ & $\langle\ln\mathcal{L}\rangle$ & $H$\\
\hline
XO4a$^{\rm L}$ & 19 & 81.88 & 0.00 & 20.87 & 215.7 & ${27.1\pm0.2}$\\
XPHM-ST$^{\rm L}$ & 19 & 81.74 & $-0.14$ & 20.83 & 214.9 & ${26.7\pm0.2}$\\
XPNR$^{\rm L}$ & 19 & 81.65 & $-0.23$ & 20.86 & 215.6 & ${27.6\pm0.2}$\\
NRSur7dq4$^{\rm L}$ & 19 & 81.61 & $-0.27$ & 20.86 & 215.4 & ${27.5\pm0.2}$\\
XO4a & 15 & 81.35 & $-0.53$ & 20.85 & 215.3 & ${28.0\pm0.2}$\\
XPNR & 15 & 79.16 & $-2.72$ & 20.61 & 210.2 & ${28.0\pm0.2}$\\
NRSur7dq4 & 15 & 78.57 & $-3.31$ & 20.70 & 212.2 & ${31.3\pm0.2}$\\
XPHM-ST & 15 & 77.00 & $-4.88$ & 20.58 & 209.4 & ${32.1\pm0.2}$\\
Glitch model & 7 & 56.29 & $-25.59$ & 17.57 & 152.0 & ${22.4\pm0.2}$\\
Power law & 7 & 53.59 & $-28.29$ & 17.33 & 147.6 & ${24.2\pm0.2}$\\
\end{tabular}
\end{ruledtabular}
\end{table}

Table~\ref{tab:event} lists the ten hypotheses. All compact-binary models are favored over noise, with $\log_{10}\mathcal{B}_{\rm s/n}\simeq77$--$82$, and over the two burst models by at least 20.7 decades. The burst models fit the merger frequency of the signal, as the cosmic-string models of Ref.~\cite{Cuceu:2025fzi} do. The power-law cutoff is at $f_{\rm high}=65.0^{+0.5}_{-1.1}$~Hz, and the sinusoid has a frequency $f_0=53.8^{+0.8}_{-0.7}$~Hz and a damping time $\tau=46^{+8}_{-7}$~ms. Both recover a network SNR of $\simeq17.5$, and the compact-binary models recover $\simeq20.7$. 

Among the unlensed models, XO4a has the highest evidence and XPHM-ST the lowest: $\log_{10}\mathcal{B}^{\rm XO4a}_{\rm XPHM\text{-}ST}=4.35$ separates two waveform models of the same physical hypothesis. Each waveform model favors its lensed version, with $\log_{10}\mathcal{B}^{\rm L}_{\rm U}=0.53$ (XO4a), 2.50 (XPNR), 3.03 (NRSur7dq4), and 4.74 (XPHM-ST). The lensing Bayes factor grows with the evidence deficit of the unlensed fit and is smallest for XO4a, the unlensed model with the highest evidence. Ref.~\cite{LIGOScientific:2025cwb} found the same trend for a point-mass lens, with $\log_{10}\mathcal{B}^{\rm L}_{\rm U}=0.1$ for XO4a and 3.8 for XPHM-ST. Lensing reduces the spread of $\log_{10}\mathcal{B}_{\rm s/n}$ across the four waveform models from 4.35 to 0.27 decades. The lensed hypothesis recovers most of the evidence deficit of each waveform model.

\begin{figure}
\includegraphics[width=\columnwidth]{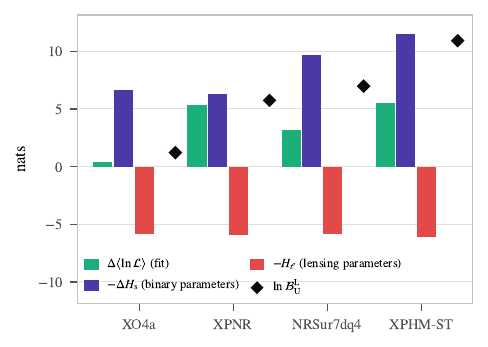}
\caption{Split of the lensing Bayes factor of each waveform model for GW231123 [Eq.~\eqref{eq:info}], in nats: gain in fit, i.e., in mean log-likelihood (green), information released by the binary parameters (purple), and information carried by the lensing parameters (red). Diamonds mark $\ln\mathcal{B}^{\rm L}_{\rm U}$, the sum of the three terms.}
\label{fig:budget}
\end{figure}

Figure~\ref{fig:budget} splits each lensing Bayes factor into fit and Occam terms [Eq.~\eqref{eq:info}] for GW231123. Lensing improves the fit by $\Delta\langle\ln\mathcal{L}\rangle=0.4$--$5.5$~nats, most for the waveform models that fit the signal worst. It also lowers the information gained from the data by 0.4--5.4~nats: the second image loosens the constraints on the binary parameters, which release more information than the lensing parameters carry. Both terms thus favor lensing for every waveform model, although for XPNR the Occam term, $0.4\pm0.3$~nats, is consistent with zero. The fitted second image is weak and arrives a few tens of milliseconds after the first, so over the few cycles of the signal it acts as a smooth correction to the amplitude and phase of the waveform. Genuine lensing would produce these signatures, and so would a flexible correction to an inaccurate waveform model. We use simulated signals to tell the two apart.

\begin{table}
\caption{Evidence results of 68 simulated signals drawn from the posteriors of four hypotheses fitted to GW231123 (rows): unlensed XO4a, unlensed XPHM-ST, lensed XO4a (XO4a$^{\rm L}$), and lensed XPHM-ST (XPHM-ST$^{\rm L}$). Each signal is analyzed with the same four hypotheses. The first three columns give the number of signals for which the true hypothesis has the highest evidence: without noise, in Gaussian noise, and in real noise. The last two give the median of $\Delta\log_{10}\mathcal{Z}_{\rm true}=\log_{10}\mathcal{Z}_{\rm true}-\max\log_{10}\mathcal{Z}$, and the hypothesis that most often has the highest evidence, with the number of signals. The last row gives the hypothesis with the highest evidence for GW231123.}
\label{tab:inj}
\renewcommand{\arraystretch}{1.15}
\footnotesize
\begin{ruledtabular}
\begin{tabular}{lccccc}
 & \multicolumn{3}{c}{True has highest $\mathcal{Z}$} & Median & Most often\\
\cline{2-4}
Simulated & No noise & Gaussian & Real & $\Delta\log_{10}\mathcal{Z}_{\rm true}$ & highest\\
\hline
XO4a & 1/1 & 4/12 & 2/4 & $-0.04$ & XO4a$^{\rm L}$ (9)\\
XPHM-ST & 0/1 & 0/12 & 0/4 & $-1.25$ & XO4a$^{\rm L}$ (9)\\
XO4a$^{\rm L}$ & 0/1 & 7/12 & 3/4 & $0.00$ & XO4a$^{\rm L}$ (10)\\
XPHM-ST$^{\rm L}$ & 0/1 & 4/12 & 1/4 & $-0.07$ & XO4a$^{\rm L}$ (12)\\
\hline
All & 1/4 & 15/48 & 6/16 & & XO4a$^{\rm L}$ (40)\\
GW231123 & & & & & XO4a$^{\rm L}$\\
\end{tabular}
\end{ruledtabular}
\end{table}

Table~\ref{tab:inj} summarizes the evidence values obtained for the 68 simulated signals. The true hypothesis has the highest evidence for only 22 of them: one of the four without noise, 15 of the 48 in Gaussian noise, and 6 of the 16 in real noise. Most misses are ties: for 47 signals, the true hypothesis and the best alternative are within half a decade, a difference ``not worth more than a bare mention''~\cite{kass1995bayes}. The true hypothesis is disfavored by more than this for 17 signals, 14 of them unlensed XPHM-ST, which never has the highest evidence for its own signals. Whatever the true hypothesis, lensed XO4a has the highest evidence most often, for 40 of the 68 signals, as for GW231123. Real noise behaves as Gaussian noise: for the real-noise signals, 44 of the 48 Bayes factors (two lensing and one between waveform models for each signal) lie within the range found in Gaussian noise for the same hypothesis.

\begin{figure}
\includegraphics[width=\columnwidth]{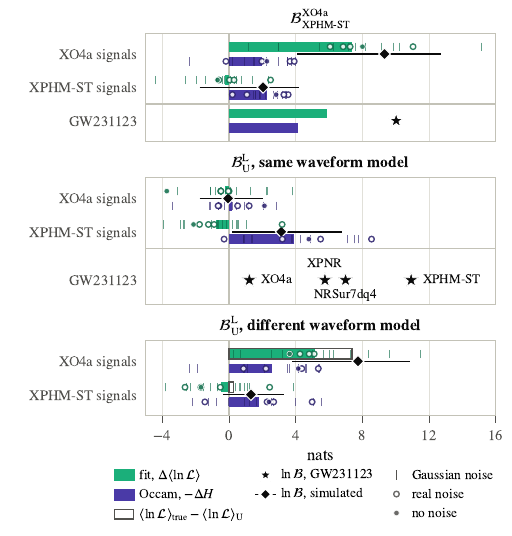}
\caption{Fit and Occam terms of $\ln\mathcal{B}=\Delta\langle\ln\mathcal{L}\rangle-\Delta H$ [Eq.~\eqref{eq:info}] for the simulated unlensed signals and for GW231123. Top: unlensed XO4a against unlensed XPHM-ST. Middle and bottom: lensed against unlensed, analyzed with the waveform model that generated the signal (middle) or with the other one (bottom). Bars are means over the 17 signals of each kind, and diamonds and black lines give the mean and range of $\ln\mathcal{B}$. Individual signals are shown as ticks (Gaussian noise), open circles (real noise), and filled circles (no noise). Outlines mark how much worse the unlensed analysis model fits than the true one, $\langle\ln\mathcal{L}\rangle_{\rm true}-\langle\ln\mathcal{L}\rangle_{\rm U}$. Stars mark $\ln\mathcal{B}$ for GW231123, with its two terms (top) and for the four waveform models (middle).}
\label{fig:inj}
\end{figure}

Figure~\ref{fig:inj} shows how these outcomes arise, by splitting the Bayes factors of the unlensed signals into data fit and Occam terms; we quote means and counts over all signals, as plotted. The fit recognizes XO4a signals: XPHM-ST fits them 7.4~nats worse than XO4a, and XO4a is favored for all 17 of them. It does not recognize XPHM-ST signals, which XO4a fits about as well as XPHM-ST, by either measure of the fit. The evidence then follows the Occam term: XO4a gains 2.3~nats less information from these signals and is favored for 15 of the 17, by $\log_{10}\mathcal{B}^{\rm XO4a}_{\rm XPHM\text{-}ST}=0.92$ for the one without noise. Ref.~\cite{Bini:2026kwz} found the same for NRSur7dq4 signals: XO4a is favored by $\log_{10}\mathcal{B}^{\rm XO4a}_{\rm NRSur7dq4}=1.5$ without noise, although NRSur7dq4 has the higher maximum likelihood, and in 16 of 20 Gaussian-noise realizations. They concluded that ``Bayes factors should not be used to resolve waveform systematics.'' More generally, the Occam term outweighs the fit for 11 of the 17 signals whose true hypothesis is clearly disfavored.

Lensing Bayes factors of unlensed signals follow the same two routes to their values: data fit versus Occam. With the true waveform model used for analysis (Fig.~\ref{fig:inj}, middle), lensing does not improve the fit on average, yet XPHM-ST favors lensing for all 17 of its signals, up to $\log_{10}\mathcal{B}^{\rm L}_{\rm U}=2.94$, mostly through the Occam term. Of the 26 unlensed signals that favor lensing with their true model, only five do so mainly through the fit, by fitting noise. When analyzing with the other waveform model (bottom), lensing absorbs most of the mismatch: on XO4a signals, lensed XPHM-ST recovers 5.2 of the 7.4~nats that unlensed XPHM-ST lacks, and reaches $\log_{10}\mathcal{B}^{\rm L}_{\rm U}=4.72$. XO4a, which fits XPHM-ST signals almost as well, has little to absorb.

GW231123 shows the second pattern: gaining in Bayes factor through mitigating the fit deficit of using the wrong waveform model. With XPHM-ST, lensing improves its fit by 5.5~nats, more than for any XPHM-ST signal analyzed with XPHM-ST (at most 3.2~nats) and about as much as for XO4a signals analyzed with XPHM-ST. With XPNR, NRSur7dq4, and XPHM-ST alike, the lensed fit comes within 0.4~nats of that of unlensed XO4a (Table~\ref{tab:event}). For XPHM-ST, $\log_{10}\mathcal{B}^{\rm L}_{\rm U}=4.74$ is level with the largest value for unlensed XO4a signals analyzed with XPHM-ST, 4.72, and for XO4a, 0.53 lies within the range of unlensed XO4a signals. The lensing Bayes factor of GW231123 thus measures how poorly each unlensed waveform model fits the signal: it is dominated by waveform systematics. The evidence cannot tell which waveform model is adequate (Fig.~\ref{fig:inj}, top), although the fit suggests that XPHM-ST is not: it fits GW231123 5.9~nats worse than XO4a, more than for any XPHM-ST signal (at most 2.5~nats) and close to the 7.4~nats of XO4a signals. With only 17 signals per hypothesis, these comparisons are not a calibration.

Lastly, we note that lensing analyses of GW231123 find lower masses and spins under the lensed hypothesis and present them as support for lensing~\cite{Goyal:2025eqo}. In our work, lensing likewise lowered the inferred spin of the heavier black hole for all 34 unlensed signals analyzed with XO4a, by a median of 0.25, as for GW231123 (from 0.92 to 0.68), because the second image loosens the spin constraint. A lower spin is therefore a property of the lensed model, not evidence for lensing. Without noise, Ref.~\cite{Hegde:2026anm} also concluded that extra degrees of freedom due to lensing reduce waveform evidence range while providing support for lensing. Additionally, we note that including noise, Gaussian or real, acts in support of the more flexible model due to the extra degrees of freedom. 

\section{Conclusions}
Every waveform model we tested favors a lensed interpretation of GW231123, and lensing nearly removes the evidence differences between waveform models. Simulated signals show why. Even without lensing, the hypothesis with the highest evidence is often not the true one, because the Occam penalty can outweigh the fit, as found for waveform models~\cite{Bini:2026kwz}. Unlensed signals analyzed with the waveform model that generated them favor lensing by up to $\log_{10}\mathcal{B}^{\rm L}_{\rm U}\simeq3$, mostly through this Occam penalty. Analyzed with another waveform model, they reach $\log_{10}\mathcal{B}^{\rm L}_{\rm U}\simeq4.7$, the value of GW231123, because the lensed model absorbs the mismatch between waveform models. GW231123 behaves in this second way: its lensing Bayes factor is dominated by waveform systematics. The lower spins inferred under lensing also appear for unlensed signals. These results hold without noise, in Gaussian noise, and in the real detector noise around the event.

Other extensions of the quasicircular binary behave the same way. Overlapping signals~\cite{Hu:2025lhv}, eccentricity~\cite{Wang:2026lur}, and birefringence~\cite{Liu:2026wor} are each favored by $\log_{10}\mathcal{B}\sim1$--$5$ for some waveform model, as is lensing~\cite{LIGOScientific:2025cwb,Goyal:2025eqo,Shan:2025dcd,Chan:2025kyu}, and the waveform model that favors an extension changes with the extension: lensing is largest with XPHM-ST~\cite{LIGOScientific:2025cwb,Goyal:2025eqo,Shan:2025dcd}, while birefringence is favored only with XO4a~\cite{Liu:2026wor}. Bayes factors between waveform models are as large, $\simeq4$ decades here and $\simeq5$ between TEOBResumS-Dal\'{\i} and NRSur7dq4 in~\cite{Jan:2025zcm}, and analysis settings change them too~\cite{Jan:2025zcm,Wang:2026lur}.

A lensing Bayes factor of this size indicates lensing only when compared with what unlensed signals produce~\cite{Caliskan:2022wbh}, and only if the waveform model describes the signal, which the evidence itself cannot establish. Simulations generated and analyzed with a single waveform model miss the dominant effect found here (see also Refs.~\cite{Shan:2025dcd,Chan:2025kyu}). Before interpreting a Bayes factor, we recommend testing whether each compared model can reproduce the data, e.g., with posterior predictive checks~\cite{gelman1996posterior,Romero-Shaw:2022ctb} as implemented in \Basilic~\cite{Cuceu:2026ggj}. 

GW231123 has few cycles and waveform models that disagree. These circumstances are where checks matter most, as such events become increasingly likely in the future. For GW231123, the data show that more flexible models, lensing among them, improve the fit to the signal. They do not support that the source is lensed.

\begin{acknowledgments}

This research has made use of data~\cite{LIGOScientific:2025snk} or software obtained from the Gravitational Wave Open Science Center (gwosc.org), a service of the LIGO Scientific Collaboration, the Virgo Collaboration, and KAGRA. This material is based upon work supported by NSF's LIGO Laboratory which is a major facility fully funded by the National Science Foundation, as well as the Science and Technology Facilities Council (STFC) of the United Kingdom, the Max-Planck-Society (MPS), and the State of Niedersachsen/Germany for support of the construction of Advanced LIGO and construction and operation of the GEO600 detector. Additional support for Advanced LIGO was provided by the Australian Research Council. Virgo is funded, through the European Gravitational Observatory (EGO), by the French Centre National de Recherche Scientifique (CNRS), the Italian Istituto Nazionale di Fisica Nucleare (INFN) and the Dutch Nikhef, with contributions by institutions from Belgium, Germany, Greece, Hungary, Ireland, Japan, Monaco, Poland, Portugal, Spain. KAGRA is supported by Ministry of Education, Culture, Sports, Science and Technology (MEXT), Japan Society for the Promotion of Science (JSPS) in Japan; National Research Foundation (NRF) and Ministry of Science and ICT (MSIT) in Korea; Academia Sinica (AS) and National Science and Technology Council (NSTC) in Taiwan.

The authors are grateful for computational resources provided by the LIGO Laboratory and supported by the National Science Foundation Grants PHY-0757058 and PHY-0823459. This work was supported with financial support from the French Agence Nationale de la Recherche. This material is based upon work supported by NSF’s LIGO Laboratory which is a major facility fully funded by the National Science Foundation. This manuscript was assigned LIGO-Document number P2600634.
\end{acknowledgments}

\bibliography{references}

\end{document}